\documentclass[12pt,onecolumn,prb,aps,showpacs]{revtex4}
\usepackage{graphics}
\usepackage{dcolumn}
\usepackage{bm}
\usepackage{amsmath}
\usepackage{amssymb}
\usepackage{epsfig}
\usepackage{pictex}
\begin{document}
\title{Correlation effects in the electronic structure of the Ni-based superconducting $KNi_2S_2$}
\author{Feng Lu$^{1}$,
        Wei-Hua Wang$^{2}$, Xinjian Xie$^{3}$,
    and Fu-Chun Zhang$^{1,4}$} \affiliation{\it
 $^1$  Department of Physics, and Center of Theoretical and Computational Physics, The University of Hong Kong, Hong Kong, China  \\
\it $^2$ Department of Electronics and Tianjin Key laboratory of Photo-Electronic Thin film Device and Technology, Nankai University, Tianjin, China  \\
\it $^3$ School of materials and engineering, Hebei University of Technology,Tianjin 300130, People¡¯s Republic of China
\it $^4$ Department of Physics, Zhejiang University, Hangzhou, China}
\date{today}

\begin{abstract}

The LDA plus Gutzwiller variational method is used to investigate the ground-state physical properties
of the newly discovered superconducting $KNi_2S_2$.
Five Ni-3d Wannier-orbital basis are constructed by the density-functional theory,
to combine with local Coulomb interaction to describe the normal state electronic structure of Ni-based superconductor.
The band structure and the mass enhanced are studied based on
a multiorbital Hubbard model by using Gutzwiller approximation method.
Our results indicate that the correlation effects
lead to the mass enhancement of $KNi_2S_2$.
Different from the band structure calculated from the LDA results,
there are three energy bands across the Fermi level along the X-Z line
due to the existence of the correlation effects, which induces a very complicated
Fermi surface along the X-Z line.
We have also investigated the variation of the quasi-particle weight factor with the hole or electron doping
and found that the mass enhancement character has been maintained with the doping.
\end{abstract}


\maketitle

\section{Introduction}
The discovery of the second class of high-transition temperature superconductors in LaFeAsO$_{1-x}$F$_x$\cite{Kamihara-1} has stimulated great
interest in the study of iron pnictides.
Over the past five years, many iron-based compounds were reported to show superconductivity after doping or under high
pressures\cite{Kamihara-1,Chen-G-F-1,Lei-H-1,C-1};
and the superconducting (SC) transition temperature
exceeds 50K in the $RFeAsO_{1-x}F_x$($R$: Rare earth atoms)\cite{Wug-1}.
Recently, one family $AFe_2Se_2 (A = K, Rb$, or $Tl)$ compounds with the $ThCr_2Si_2$-type crystal structure,
such as $K_yFe_2Se_2$\cite{Guo-Jin-1} and $(Tl,K)_yFe_{2-x}Se_2$ \cite{Fang-Wang-1,Wang-Dong-1},
have been found to become superconductors at about 30 K.
Different from other crystal structures of the iron-based superconductor,
these compounds $AFe_2Se_2 (A = K, Rb$, or $Tl)$ have many unique physical properties,
such as absence of hole-like Fermi surface (FS)in $K_{0.8}Fe_{1.7}Se_2$\cite{Qian-Wang-1}
and the intrinsic $\sqrt{5} \times \sqrt{5}$ Fe vacancies superstructure in $K_{0.8}Fe_{1.6}Se_2$
\cite{Wei-Bao-1,Wei-Bao-2,Wei-Bao-3}.
Thus, searching for superconductors with the $ThCr_2Si_2$-type crystal structure
has become one of the interesting topics in condensed matter physics\cite{Lee-P-A-1,C-2}.

%
\begin{figure}[tp]
\begin {center}
\vglue -0.6cm \scalebox{1.150}[1.15]
{\epsfig{figure=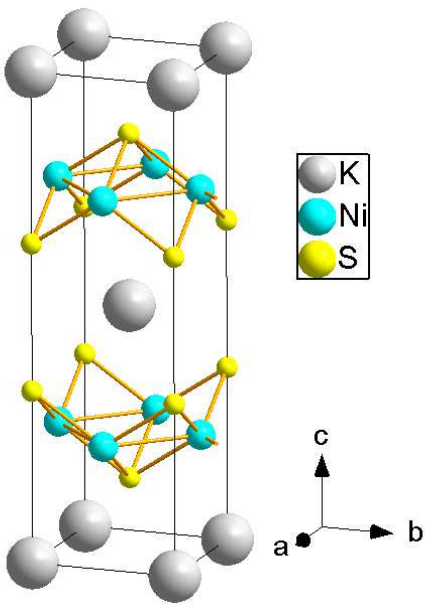,width=12.cm,angle=0}} \caption{
(Color online) $KNi_2S_2$ with the $ThCr_2Si_2$-type structure.}
\label{fig:phase-diagram-1}
\end {center}
\end{figure}
Very recently, a novel group of superconductors with $ThCr_2Si_2$-type
crystal structure, $KNi_2A_2 (A=S,Se)$, have been discovered\cite{Neilson-1,Neilson-2}.
The experiments show that the Ni-based superconductor exhibits a rich and unusual
electronic and structural phase diagram\cite{Neilson-1,Neilson-2}.
For example, based on high-resolution synchrotron X-ray diffraction and time-of-flight neutron scattering experiments,
James et al. have found that the mixed valence compound $KNi_2S_2$ displays a number of highly unusual structural transitions and presents the charge density wave fluctuations\cite{Neilson-2}.
%
The specific heat measurement reveals that $KNi_2S_2$ exhibits the correlated behavior
at low-temperature with an enhanced effective electron mass, $m^* =11m_b$ to $24m_b$.
Below $T_c=0.46 K$, $KNi_2S_2$ becomes superconducting without the partial substitution of K or Ni deficient\cite{Neilson-2}.
As discussed above,  the Ni-based superconductor $KNi_2S_2$ has a similar crystal structure as the iron-based superconductor $AFe_2S_2$ compound. However, there are also many differences between these two kinds of compounds.
For example, the localized magnetism is absent in the Ni-based superconductor $KNi_2S_2$\cite{Neilson-2}.
Thus, it is critical to understand the electronic structures of the parent $ANi_2S_2$ compounds at first
in order to investigate the mechanism of superconductivity in this material.

In this paper, we have performed the first-principle electronic structure calculations on $KNi_2S_2$.
To overcome the problem of dealing with Coulomb interaction in the local density approximation (LDA) calculation,
we use LDA plus the Gutzwiller variational approach to investigate this typical correlated material.
This method has proved to be effective in dealing with Coulomb interaction and
has been widely applied to strongly correlated systems\cite{Deng_Wang_1,Wang_Qian-2}.
The calculation shows that $KNi_2S_2$ is a kind of multi-orbital system with
five Ni-3d orbital near the Fermi level.
Our results indicate that the correlation effects of the 3d local electrons
lead to mass enhancement of $KNi_2S_2$.
This behavior is always present with the hole or electron doping,
which is well agreement with the experimental results\cite{Neilson-2}.
The correlation effect is not only presented by the narrowing of the bandwidth,
but also by the change of the FS near the high symmetry line.
In the LDA calculation, there are two bands across the X-Z line.
Actually, three electron bands across the Fermi level along the X-Z line,
which induce a very complicated FS structure along the X-Z line.
The relationship among the quasi-particle weight factors of the different orbital is also discussed in this paper.

\section{Model and method}
\label{secmodel}

The first-principles calculations were performed within LDA using the plane wave basis method\cite{P-Giannozzi-1}.
The generalized gradient approximation (GGA) was adopted for the exchange-correlation potential of the density function theory (DFT)\cite{Perdew-1}.
We used the ultrasoft pseudopotentials to model the electron-ion interactions\cite{D-Vanderbilt-1}.
The kinetic energy cutoff and the charge density cutoff were taken as 4800 and 6400 eV, respectively.
A mesh of $18 \times 18 \times 9$ k points were used to perform the Brillouin zone integration.
$KNi_2S_2$ has $ThCr_2Si_2$-type structure (space group I4$/$mmm, ($\#139$, Z=1))
and each unit cell contains two formula units, as shown in Fig.1.
In our calculation, the structure parameters optimized by the energy minimization,
a=3.7792\AA, c=12.7139\AA,  and $z(S)$=0.35, were adopted,
which are in good agreement with the experimental values
a= 3.80899\AA, c=12.67609\AA,  and $z(S)$=0.34642 determined by synchrotron X-ray powder diffraction\cite{Neilson-2}.
Since the LDA can not describe the Coulomb interaction adequately,
in order to take into account more explicitly the correlated effects of the d electrons, we
have performed  Gutzwiller variational approach to deal with the Coulomb interaction.

\begin{figure}
\includegraphics[clip,scale=0.32]{Fig-2a.eps}
\includegraphics[clip,scale=0.33]{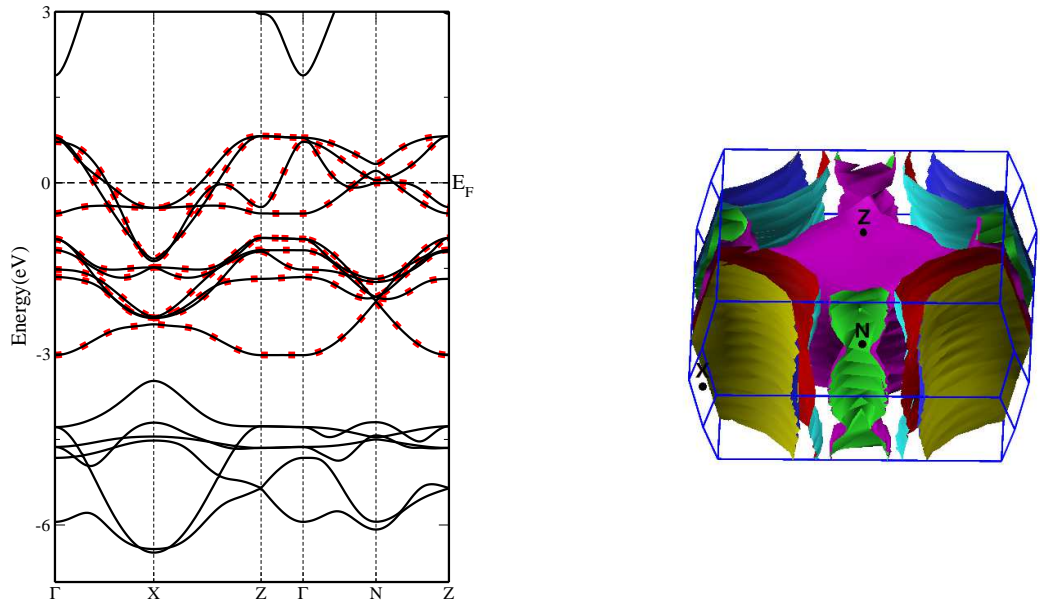}
\caption{(Color online) (a) Energy band structure of $KNi_2S_2$ by LDA(black line) and Wannier function(dotted line) ;
(b) The Fermi surface of $KNi_2S_2$ by LDA calculation.
}
\end{figure}

\section{Results and discussions}
\label{secmodel}

We first present the electronic structure of $KNi_2S_2$ by LDA calculation.
The corresponding band structures and FS are given in Fig. 2a and Fig. 2b, respectively.
Similar to the iron-based superconductor\cite{Mazin-I-I-1,Singh-D-1,Boeri-L-1,C-Cao-1,F-Ma-1,lufeng-1},
there are several bands across the Fermi level,
which indicates that the $KNi_2S_2$ is a kind of multi-orbital superconducting system.
The multi-orbital character can also be seen from the complicated FS sheets.
As shown in Fig.2 b,
there is a large hole-like FS around the $\Gamma$ point,
two 2D-cylinder-like ones around the corners of the Brillouin zone and
a very complicated pocket at the middle of the four boundaries of the Brillouin zone.
As mentioned above, the structure of the FS is very complicated
due to the multi-orbital characters,
which implies that the $KNi_2S_2$ should be a kind of multiple-gap superconductors similar to
the iron based superconductor, in contrast to the single-orbital copper-based superconductors.

%
\begin{figure}[tp]
\begin {center}
\vglue -0.6cm \scalebox{1.150}[1.15]
{\epsfig{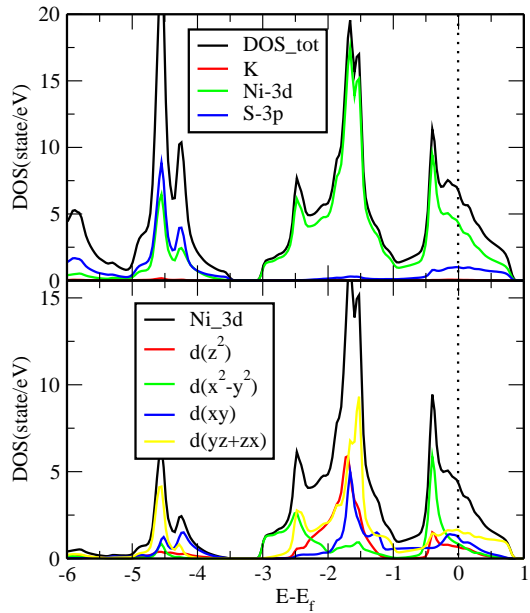}} \caption{
(Color online) (a) DOS and PDOS of $KNi_2S_2$ .
(b) PDOS of $KNi_2S_2$ for the five Fe-3d orbitals. }
\label{fig:phase-diagram-3}
\end {center}
\end{figure}

To investigate the orbital characters near the FS,
the corresponding density of states (DOS)
and the projecting the density of states (PDOS) of $KNi_2S_2$ are plotted in Fig.3.
The basic shape of the DOS and PDOS is very
similar to that of $AFe_2Se_2$\cite{Yan_Miao_3}.
However, the position of the Fermi energy is higher due to the more electron count.
As shown by the DOS in Fig.3, the total DOS
near the $E_f$ is manly contributed by the Ni-3d orbitals,
which mix with the S-3p orbitals near the $E_f$.
In detail, the PDOS of the Ni-3d orbitals consist of the DOS from -3 eV to 1 eV around the $E_f$,
and at the same time there is some contribution from S-3p orbitals.
Evidently, the contribution of S-3p orbitals is smaller
than that from the Ni-3d orbitals.
The value of DOS at the Fermi energy N($E_f$) is about 6.9 states
per eV per formula unit both spins.

\begin{figure}
\includegraphics[clip,scale=0.6]{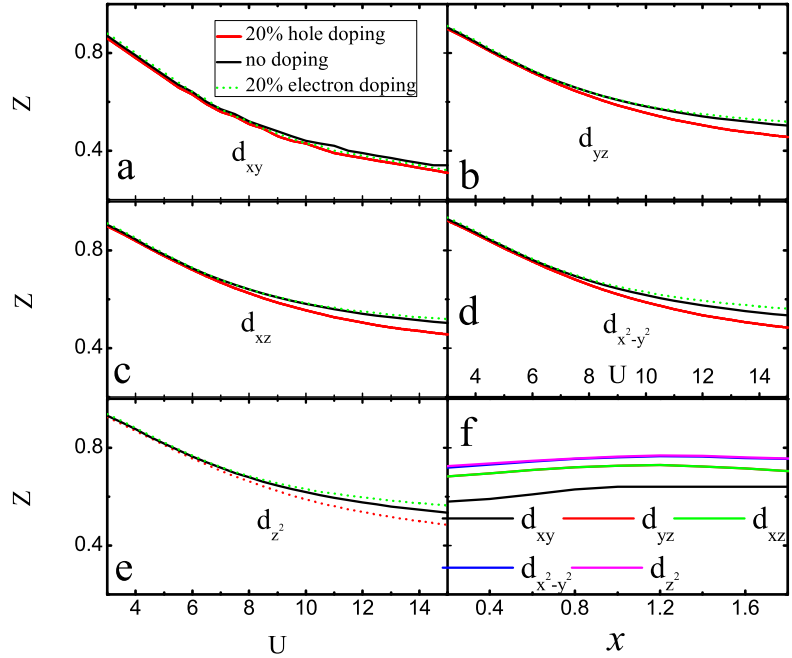}

\caption{(Color online) The quasi-particle weight factor Z as a function of Coulomb interaction U at J=0.8eV.
(a) The quasi-particle weight factor Z for $d_{xy}$ orbital;
(b) The quasi-particle weight factor Z for $d_{yz}$ orbital;
(c) The quasi-particle weight factor Z for $d_{xz}$ orbital;
(d) The quasi-particle weight factor Z for $d_{x^2-y^2}$ orbital;
(e) The quasi-particle weight factor Z for $d_{z^2}$ orbital;
(f) The quasi-particle weight factor Z as a function of doping $x$ for $K_xNi_2S_2$ at U=8 eV and J=0.8eV.
}
\end{figure}

As we know, both LDA and GGA
fail to depict the strongly correlated electron systems
due to the insufficient treatment of electron correlation.
In the last 20 years, a lot of new methods and new computational
tools have been developed to improve the calculation of
strongly correlated electron materials, such as the LDA+U\cite{ Anisimov_Zaanen-1},
LDA+DMFT(dynamical mean-field theory) \cite{Kotliar_Savrasov-1}
and LDA+G (Gutzwiller method)\cite{Deng_Wang_1}.
Among all these methods, LDA+G is thought as
one that is practically efficient and can
capture the key feature of the correlation effect as well
in the ground-state studies.
The Gutzwiller
variational approach\cite{Gutzwiller-1} was firstly introduced by  Gutzwiller
and has been proved to be quite efficient
and accurate for the ground-state physical properties studies,
such as the the Mott transition, ferromagnetism,
and superconductivity\cite{Deng_Wang_1, Gutzwiller-1, Vollhardt-1, Brinkman-1, Zhan_Gros_1}.
In our calculation, we also used the LDA+G framework to treat the correlation effects between
the localized d electrons.
In other words, we firstly construct the multi-orbital Hubbard model and then solve this model by the
Gutzwiller variational approach (GW).
Firstly, the Wannier function is constructed by
the maximally localized Wannier functions (MLWFs) approach\cite{Nicola_Arash-1},
as shown by the dotted line in Fig.2 a.
Secondly, combining with the Coulomb interaction, the multi-orbital model can effectively describe
the multi-orbital characters of $KNi_2S_2$.
Then the GW method is used to deal with this many-body model and the physical properties in this system
\cite{Deng_Wang_1, Gutzwiller-1, Vollhardt-1, Brinkman-1, Zhan_Gros_1}.

We start from this five-orbital Hubbard model,
\begin{eqnarray}
   H_{Tot} &=& H_{wannier}+H_{int}
          \label{eq:model},
\end{eqnarray}
where $H_{wannier}$ describes the on-site energies and hopping integrals
of electrons between the Ni-3d
Wannier orbitals;
and $H_{int}$ is the local interaction term for the Coulomb interaction
between the d electron described by the parameters $U$, $U^{\prime}$, $J$.
The parameters $U$, $U^{\prime}$, and $J$ denote the
intra-orbital Coulomb interaction, inter-orbital Coulomb interaction and Hund's coupling.
In what follows, considering the realistic
wavefunctions of 3d-orbitals \cite{Castellani} and the spin
rotational symmetry, we adopt the relationships $U=U'+2J$.

The true ground state of $H_{Tot}$ is described by the Gutzwiller wave function $|\Psi_G>$,
which is constructed by a many-particle projection operator $\hat{P}_{G}$
on the uncorrelated wave function $|\Psi_0>$
\begin{eqnarray}
|\Psi> & = & \hat{P}_{G}|\Psi_{0}>=\prod_{i}\hat{P}_{i}|\Psi_{0}>
\end{eqnarray}
with Gutzwiller projector operator
\begin{eqnarray}
\hat{P}_{i} & = & \sum_{\Gamma}\lambda_{\Gamma}|\Gamma><\Gamma|
\end{eqnarray}
where $\lambda_{\Gamma}$ is the variational parameter for the
i-th site with the atomic eigenstates $|\Gamma>$, which is
determined by minimizing the ground state total energy.
And the renormalization of kinetic energy can be described by the bandwidth renormalization or quasi-particle weight factor Z,
as defined in Ref. 17.

Generally speaking, the essence of the correlation effects under Gutzwiller approach is to suppress the double occupancy and to renormalize the kinetic energy.
Specifically, the Coulomb interaction U will enhance the orbital polarization; however, the interorbital Hund¡¯s coupling J favors even distribution of electrons among five orbitals.
To understand the role of Coumlomb interaction and the change of the band structure in the presence of interactions,
the value of the quasi-particle weight factor Z of the kinetic energy
as a function of U at fixed J=0.8 eV is shown in Fig. 4.
As shown by the solid line in Figs. 4(a),
the Z factor decreases with the increase of the Coulomb interaction U,
due to the suppression of the double occupancy by U.
Because of the valence state of $Ni^{1.5+}$,
each five Ni-3d orbitals have 8.5 electrons and are non-integer filling.
Thus, the metal insulator transition does not occur in this system and $KNi_2S_2$ is a multi-orbital metal system, which is agreement with the experiment\cite{Neilson-2}.
In the iron-based or the copper-based superconductors, different doping with holes ($p$ type) or electrons ($n$ type) has important influence on the superconducting transition temperature,
which lead to a rich phase diagram of these high temperature superconductors.
For example, superconductivity only occurs in a doping range, which indicates that electron doping or hole doping may affect the electronic properties in these materials.
To study the doping effect on the physical properties of $KNi_2S_2$ ,
we investigate the quasi-particle weight factor Z at different doping situations,
as shown in Fig. 4(a-e).
The dotted line shows Z at the $20\%$ hole doping,
the black solid line shows Z without doping and
the dashed line shows Z at the $20\%$ electron doping, respectively.
When the Coulomb interaction U is not too large,
the electron or hole doping almost do not affect the quasi-particle weight factor Z.
But, the difference of the quasi-particle weight factor Z with different doping become
more and more obvious with the increase of the U in the large range;
and the value of Z is also dependent on the orbital degree of freedom due to the orbital anisotropy.
For example, the doping effect is very weak for the $d_{xy}$ orbital but important for the $d_{z^2}$ orbital
in the large U range, as shown in Fig. 4(a-e).
At fixed Coulomb interaction value U=7 eV,J=0.8 eV,
the variation of the quasi-particle weight factor Z with the different doping is also investigated and plotted in Fig. 4f.
The results show that the correlated electron character of the $KNi_2S_2$ has been maintained with the electron-like or hole-like doping for all the Ni 3-$d$ orbitals,
which is well agreement with the experimental measurements\cite{lei-1}.

%
\begin{figure}[tp]
\begin {center}
\vglue -0.6cm \scalebox{1.150}[1.15]
{\epsfig{figure=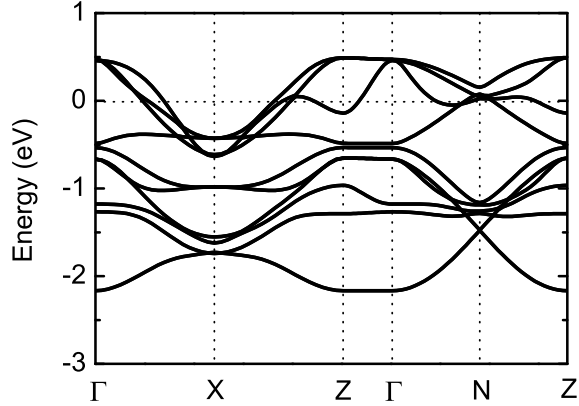,width=8.cm,angle=0}} \caption{
(Color online) Electronic band structure of $KNi_2S_2$ calculated by the Gutzwiller method.
}
\label{fig:phase-diagram-5}
\end {center}
\end{figure}

%
\begin{figure}[tp]
\begin {center}
\vglue -0.6cm \scalebox{1.150}[1.15]
{\epsfig{figure=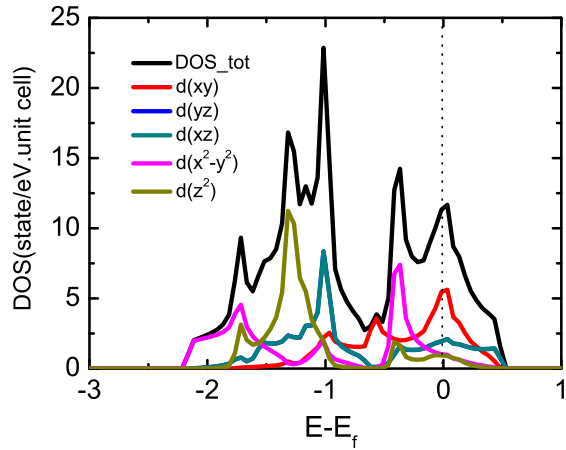,width=8.cm,angle=0}} \caption{
(Color online) DOS and PDOS of $KNi_2S_2$ calculated by the Gutzwiller method.}
\label{fig:phase-diagram-6}
\end {center}
\end{figure}

The corresponding band structures and PDOS of the Ni-based superconducting
compound $KNi_2S_2$ calculated by the GW method are presented in Fig. 5 and Fig. 6.
Since the electron occupation number is 8.5 for each Ni-3d orbitals
and the five Ni-3d orbitals are almost occupied by the electrons,
the added Coulomb interactions can not make $KNi_2S_2$ become a Mott insulator state,
which is agreement with the experiment\cite{Neilson-2}.
At a typical value of Coulomb interaction U=8 eV and J=0.8 eV \cite{Lu_Zhao-1,Stewart_Liu-1,Han_Wang-2},
the quasi-particle weight factor is about 0.5.
In other words, the overall bandwidth calculated by the GW method is smaller than that by the LDA calculation
due to the strong correlation effect among the Ni-3d orbitals electrons.
In the LDA results, the band structure near Fermi level along the X-Z line is very simple,
which causes a two FS structure along the X-Z line.
Actually, our GW calculation shows that there are three Bloch bands across the Fermi level along the X-Z line
due to the existence of the correlation effect, which would induce a very complicated FS along the X-Z line.
The corresponding electron FS
will dominate the low energy properties along the X-Z line,
which is easier to be observed by the Angle-resolved photoemission spectroscopy experiments (ARPES).
As shown above, comparing with the LDA calculation,
the GW calculation can improve the correct band narrowing and mass renormalization,
which can also be seen from the DOS calculated by the LDA+G method, as shown in the Fig.6.
The corresponding orbital character is also plotted in Fig.6.
Comparing the orbital character of Ni-based superconductor with that of the iron based superconductor\cite{Wang_Qian-2},
we can clearly find that all the Ni-3d orbitals are renormalized by the correlation effect.
We obtain the value of DOS at the Fermi energy N($E_f$) is about 11.5 states
per eV per formula unit both spins,
which corresponds to a bare specific heat coefficient, $\gamma=27$ mJ mol f.u.$^{-1}$ K$^{-2}$
based on the formula $\gamma=\pi^2k^2 N(E_f)/3$\cite{Singh_2,Beck_Claus_1}.
This $\gamma$ value is 2.5 times smaller than the experimental value $\gamma=68$ mJ mol f.u.$^{-1}$ K$^{-2}$\cite{Neilson-2},
but it is better than that of the LDA calculation, $\gamma=16.2$ mJ mol f.u.$^{-1}$.

%
\section{Conclusions}
In conclusion, we have presented the results of the electronic structure of $KNi_2S_2$ based on LDA plus Gutzwiller variational calculations.
The results show that the ground state of $KNi_2S_2$ is a strongly correlated metal and
the energy bands near Fermi level are dominated by the Ni-3d electron states.
The multi-orbital correlation effect between the Ni-3d electrons determines the main physical properties of
newly discovered Ni-based superconductor $KNi_2S_2$, such as the mass enhancement behavior and the multi-orbital characters.
Different from the LDA understanding,
the band structure has a lot of changes under correlation effect,
especially for the band structure along the X-Z high symmetry line.
In particular, the bands are narrowed
by a factor of about 2, and the five Ni-3d orbital are all almost filled, which determine the
low energy physical properties of $KNi_2S_2$.
The calculation about the electronic structure provides a reasonable
starting point for the investigations of the Ni-based superconductor.
The variation of the quasi-particle weight factor with the doping is also investigated and compared to the experiments.
Our results confirm that $KNi_2S_2$ is a multi-orbital strongly correlated system with a reduced renormalized
bandwidth by the correlation effect.

\begin{acknowledgments}
The authors thank X. Dai for helpful discussions.
FCZ thanks the
support of RGC HKU707211 and AOE/P-04/08.
\end{acknowledgments}

\end{document}